\documentclass[prb,twocolumn,showpacs,amsmath,amssymb,superscriptaddress]{revtex4}
\usepackage{graphicx}
\usepackage{bm}

\begin{document}
\title{Hidden Scaling in the Quantum Hall Metal-Insulator Transition}
\author{L. Moriconi}
\affiliation{Instituto de F\'\i sica, Universidade Federal do Rio de Janeiro, \\
C.P. 68528, 21945-970, Rio de Janeiro, RJ, Brazil}
\author{Ana L.C. Pereira}
\author{P.A. Schulz}
\affiliation{Instituto de F\'\i sica Gleb Wataghin, Universidade Estadual de Campinas, \\
C.P. 6165, 13083-970, Campinas, SP, Brazil}

\begin{abstract}
Scaling properties of the quantum Hall metal-insulator transition are severely affected by finite size effects in small systems. Surprisingly, despite the narrow spatial range where probability structure functions exhibit multifractal scaling, we clearly verify the existence of extended self-similarity -- a hidden infrared scaling phenomenon related to the peculiar form of the crossover at the onset of nonmultifractal behavior. As finite size effects get stronger for structure functions with negative orders, the parabolic approximation for the multifractal spectrum loses accuracy. However, by means of an extended self-similarity analysis, an improved evaluation of the multifractal exponents is attained for negative orders too, rendering them consistent with previous results, which rely on computations performed for considerably larger systems. \end{abstract}


\maketitle

The quantum Hall metal-insulator transition has been the focus of great interest, proving to be an ideal stage for the interplay of several approaches in the physics of condensed matter localization. \cite{girvin,jansen} In fact, while a number of field theory descriptions have been proposed, \cite{levine,zirnb,bhaseen} numerical simulations and experiments are relatively precise, \cite{huck} and reduced lattice models \cite{chalker1,kondev} amenable to theoretical treatment, have an interesting connection with the classification problem of disordered systems. \cite{zirnb2} Even though the basic physical picture of the quantum Hall metal-insulator transition was put forward around two decades ago, \cite{klitz} a lot of work has been devoted to the subject along the last years. In particular, the complete theoretical characterization of the multifractal behavior at the transition, \cite{lud} and an exact solution, likely in terms of some conformal field theory, \cite{zirnb} are so far open problems.

Our aim in this work is to call attention to a hidden scaling property of the non-interacting quantum Hall metal-insulator transition, named extended self-similarity (ESS). This kind of scaling phenomenon was found for the first time in the context of fully developed turbulence, \cite{benzi1,benzi2} where it has been a fundamental tool to extract information on the turbulent regime out of moderate Reynolds number data. After the pioneering works of Benzi et al., ESS has been verified in other classical multifractal dynamical systems, as chaotic advection, \cite{segel},
kinetic roughening, \cite{kundag} turbulent magnetohydrodynamics, \cite{basu} and strange attractors. \cite{mori} Regarding the quantum domain, a clear numerical evidence of ESS in the two-dimensional metal-insulator transition has been recently reported, \cite{mori} for a model of random Dirac fermions, which is a close variant of the quantum Hall problem. \cite{lud,chalker2}

In short words, the ESS phenomenon corresponds to a specific form of crossover for the scaling functions, that occur between small and large length scales. More concretely, let $L$ be the linear system size, $\psi(\vec x)$ be the wavefunction for some fixed energy level and $\Omega$ be a square with size $r \times r$. We define the probability structure function of order $q$ as
\begin{equation}
S_q(r)= \langle [\int_\Omega d^2 \vec x | \psi( \vec x)|^2]^q \rangle \ , \label{struct}
\end{equation}
where the average is taken over disorder realizations. The multifractal scaling law $S_q(r) \sim r^{\zeta(q)}$ supposedly holds at the metal-insulator transition. However, if finite size effects are taken into account or if the disordered potential has a finite correlation length, the scaling law is valid only through a certain range of scales. 
Either towards the smaller or larger length scales (the ``ultraviolet" and ``infrared regimes", respectively), wavefunctions will become effectively smooth, \cite{mori} leading to the trivial scaling $\zeta(q) = 2q$. As $r$ moves out from the scaling region, towards the infrared, the general meaning of ESS is just a peculiar profile of scaling, in the form \cite{comment}
\begin{equation}
S_q(r) = c_q [f_1(r)]^{\zeta(q)}[f_2(r)]^{2q} \ . \ \label{ess1}
\end{equation}
The crossover functions introduced above behave as $f_1(r) = r/L$ and $f_2(r) = 1$ in the scaling region and as $f_1(r) = 1$ and $f_2(r) = r/L$ in the infrared regime, where finite size effects become relevant. 

In the particular problem of localization, it follows from translation invariance and from definition (\ref{struct}) that $S_1(r) = (r/L)^2$ is an exact relation. Taking this result into account together with Eq. (\ref{ess1}), we find that
\begin{equation}
S_q(r)(\frac{r}{L})^{-2q} = c_q [f_1(r)]^{\zeta(q)-2q} \ . \ \label{ess2}
\end{equation}
Therefore, even if a plot of $\ln[S_q(r)]$ against $\ln(r)$ does not exhibit scaling when larger length scales are included, a plot of $\ln[S_p(r)r^{-2p}]$ against $\ln[S_q(r)r^{-2q}]$ will do: numerical data collapses on a straight line, with slope
\begin{equation}
\eta(p,q) \equiv \frac{\zeta(p)-2p}{\zeta(q) - 2q} \ . \ \label{ess-slope}
\end{equation}
As a consequence, experimental or numerical evaluations of $\eta(p,q)$ would be subject to smaller errors than the scaling exponents of probability structure functions.

In order to investigate the issue of ESS, we model the quantum Hall metal-insulator transition in a square lattice through the nearest-neighbor hamiltonian
\begin{equation}
H=\sum_i \varepsilon_i c^\dagger_i c_i + \sum_{\langle i, j \rangle} V ( e^{i \phi_{ij}} c^\dagger_i c_j
+ e^{-i \phi_{ij}} c^\dagger_j c_i) \ . \ \label{ham}
\end{equation}
Defining the lattice parameter as the unit length, the Landau gauge gives $\phi_{ij} = 0$ along the $\hat x$ direction and $\phi_{ij}= \pm 2\pi x \Phi / \Phi_0$ along the $\mp \hat y$ direction, where the magnetic flux ratio is $\Phi / \Phi_0 = Be/h$. On-site disorder is provided by the random energies $\varepsilon_i$, correlated within a length scale of order $\lambda$. In practice, they may be defined as $\varepsilon_i = 1/\pi \lambda^2 \sum_j a_j \exp(-|\vec x_i-\vec x_j|^2/\lambda^2)$, in terms of uncorrelated random variables $a_i$, with  $|a_i| \leq W/2$.

The simulations were taken for a $60 \times 60$ lattice, with periodic boundary conditions and $\Phi / \Phi_0 = 1/20$. For this choice of parameters, the linear dimension of the system is $L \simeq 34 \ell$, where $\ell = \sqrt{\hbar / eB}$ is the magnetic length. Furthermore, we set $\lambda=2$, the hopping parameter $V=0.14$, and $W/V=1.6$.
\begin{figure}[tbph]
\includegraphics[width=7.5cm, height=6.5cm]{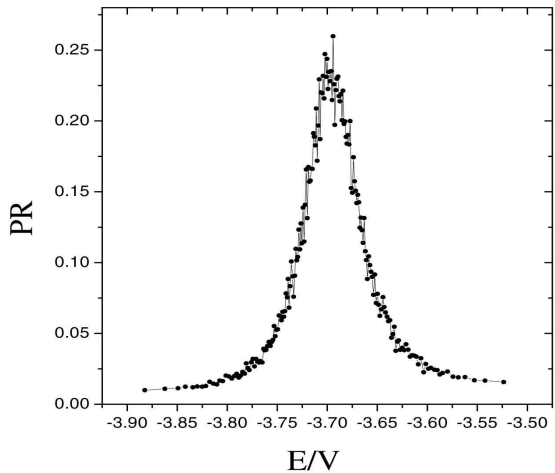}
\caption{The averaged participation ratio as a function of energy for the first Landau level. The averages are taken over 45 disorder realizations in a $60 \times 60$ lattice. The system's linear size is $L \simeq 34 \ell$.}
\label{fig1}
\end{figure}

According to the generally accepted picture, the large quantum degeneracy of Landau levels is broken due to disorder,
and delocalized states are found only around a critical energy close to each band center.
A careful determination of the critical energy, which takes into account the  levitation of extended states, \cite{ana-peter} is based on the computation of the averaged participation ratio \cite{thouless,wegner} (PR), written in terms of the wavefunction amplitude $\Psi_i$ as $PR= \langle [ N \sum_{i=1}^N | \Psi_i|^4]^{-1} \rangle$.

In our computations of the participation ratio, which are restricted to the first Landau level and shown in Fig.1, the total number of sites is $N=60 \times 60$, and the above average is taken over a set of 45 disorder realizations. For each one of them, the hamiltonian (\ref{ham}) was exactly diagonalized, producing 180 ($=60 \times 60 \times  1/20$) states per Landau level. The more localized is a state the lesser is its participation ratio.

Thus, to study the statistical properties of the delocalized wavefunctions, it is necessary to perform a selection of states around the participation ratio peak. We focused on the energy range $ -3.703 \leq E/V \leq -3.695$, which contains 8 states. We picked up, out of these states, the one with the maximum PR for each disorder realization. Had we chosen all of them for each realization, the statistical analysis would be spoiled by the several localized states that eventually appear, as a consequence of the strong fluctuations inherent to small systems. In fact, their presence can be inferred from a comparison of different selections. The selected PR's fall in the interval $0.34 \pm 0.04$. An alternative (but useless) selection which takes the minimum PR for each realization gives PR's in the interval $0.09 \pm 0.06$. On the other hand, taking all of the 8 states we get $0.23 \pm 0.1$. Of course, these three types of selection will lead to converging results in the large box limit $L \rightarrow \infty$. We assume here that our prescription is the one which yields the fastest convergence -- an hypothesis we verified a posteriori, through the computation of the multifractal exponents $\zeta(q)$.
\begin{figure}[tbph]
\includegraphics[width=7.5cm]{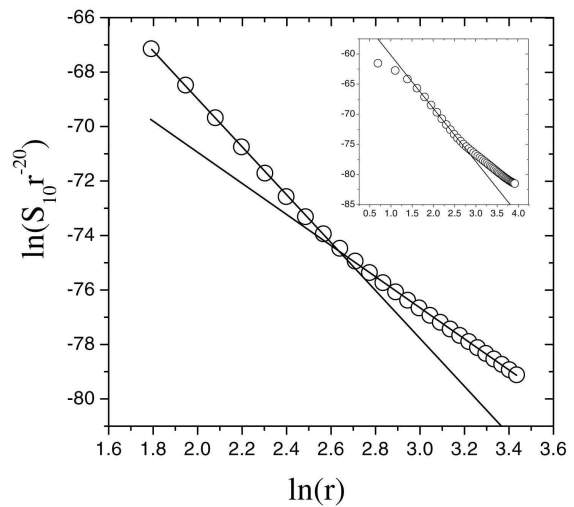}
\caption{The structure function of order q=10 reveals, when multiplied by $r^{-10}$ (circles), a scaling region ($6 \leq r \leq 14$) fit by a straight line with slope $-8.77$. The crossover to a less fluctuating regime ($14 \leq r \leq 31$) is observed, with slope $-5.72$. The inset shows the ultraviolet, scaling and the infrared regions, covering the interval $2 \leq r \leq 51$. The fitting in the inset corresponds to the steeper straight line in the large plot.}
\label{fig2}
\end{figure}

It is interesting to observe that the energy position of the participation ratio peak reaches in a fast way its asymptotic value, \cite{ana-peter} showing little additional levitation when the system's size becomes larger than $\sim 40 \times 40$, while the width and height of the PR curve are still sensitive to finite size effects.

We were able to find the scaling region of probability structure functions through an analysis of 
$S_q(r)r^{-2q}$, as shown in Fig.2. In the ultraviolet and in the infrared scales, as already discussed, 
this function becomes flatter. The scaling region, therefore, appears as an inflection between these two regimes.
This fact is shown in the inset of Fig.2 for the structure function of order $q=10$.
In general, for positive orders, scaling takes place for $6 \leq r \leq 14$ ($r$ is measured in lattice units). A crossover to the infrared regime is observed for $14 \leq r \leq 31$. In Fig.2, we establish fittings with slopes -8.77 (scaling region) and -5.72 (larger length scales). These numbers have a relative variation of $\sim 53 \%$. Similar large variations were found for other positive orders, in the interval $0 < q \leq 14$.
\begin{figure}[tbph]
\includegraphics[width=7.5cm]{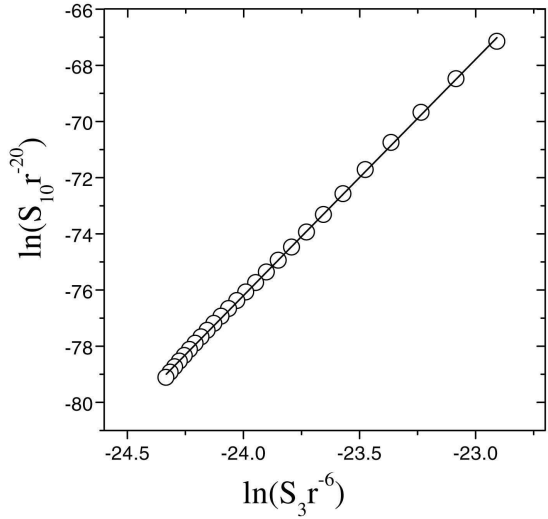}
\caption{Structure functions of different orders are compared and the phenomenon of extended self-similarity is clearly observed. The fitting, which has slope $\eta(10,3)=8.37$, approximates the whole numerical curve, and is produced taking only the data for $6 \leq r \leq 14$ (i.e., the first 9 circles from right to left).}
\label{fig3}
\end{figure}

The existence of ESS is strikingly indicated in Fig.3, the central result of our work, where we compare structure functions with orders $p=10$ and $q=3$, to obtain $\eta(10,3)$. The straight line corresponds to the fitting for $6 \leq r \leq 14$, which has slope 8.37. The fitting at larger length scales, for $14 \leq r \leq 31$, has slope 8.50. Here, the slopes have a relative $1.5 \%$ variation, which should be compared to the $53 \%$ variation obtained in Fig.2. Such higher precision evaluation is the hallmark of ESS, which was also verified for other pairs of orders $(p,q)$. A relevant question is how small or how disordered can be the system so that ESS is still meaningful. While we do not have yet a conclusive answer to this point, we have inspected simulations on a $40 \times 40$ lattice, getting results which are essentially equivalent to the ones described here.

We have checked that ESS keeps holding for negative orders, where however the parabolic approximation for the multifractal spectrum becomes inaccurate (as opposed to positive orders). The parabolic approximation simply consists of assigning a log-normal probability distribution function $\rho_r(\alpha) \sim r^{ 2-f(\alpha)}$, with $f(\alpha) \equiv 2-(\alpha - \alpha_0)^2/4(\alpha_0-2)$, to have a box probability $\int_\Omega d^2 \vec x | \psi(\vec x)|^2$ that scales as $r^\alpha$, for a given $\alpha$ parameter. \cite{jansen} The function $f(\alpha)$ is the so-called multifractal spectrum, and can be interpreted as the fractal dimension of an ``iso-$\alpha$ set". In this framework, the multifractal exponents of probability structure functions will depend solely on a single parameter $\alpha_0$. 
They are given by
\begin{equation}
\zeta(q) = 2q+q(\alpha_0-2)(1-q)  \ . \ \label{zetaq}
\end{equation}
This quadratic expression is assumed to hold as long as the multifractal spectrum is positive at the saddle-point of 
$r^{q\alpha} \rho_r( \alpha)$, i.e., we need to have $f(\bar \alpha) > 0$, with $\bar \alpha = \alpha_0 -2q(\alpha_0-2)$.
Numerical simulations, involving different types of disorder in systems with linear sizes $L \simeq 200 \ell$, yield $\alpha_0 = 2.25 \pm 0.05$ for the first Landau level. \cite{jansen,huck} More recent numerical work, performed in the Chalker-Coddington model for lattices with sizes up to $ 1280 \times 1280$, have advanced the conjecture that the parabolic approximation is in fact exact, with $\alpha_0 = 2.261 \pm 0.003$. \cite{evers}

As it can be easily checked, it is worth noting that in the parabolic approximation, we have
\begin{equation}
\eta(p,q)=\frac{p(p-1)}{q(q-1)} \ , \ \label{eta-ess}
\end{equation}
which is remarkably independent on $\alpha_0$. Computations of $\eta(p,q)$ could be used as a rigorous test of the parabolic approximation. However, it is important to point out that small errors in the multifractal exponents may lead to relatively larger errors for $\eta(p,q)$.

We show in Fig.4 the multifractal exponents $\zeta(q)$ for $-2.5 \leq q \leq 2.5$, used to analyze the parabolic approximation for the multifractal spectrum. Circles represent the direct numerical evaluations. The spatial range where ESS is now verified, viz $6 \leq  r  \leq 16$, is smaller compared to the one related to positive orders. The exponents plotted in Fig.4 (circles) correspond to linear fittings in the interval $6 \leq r \leq 10$. If the positive orders are neglected, the best fit produced from Eq. (\ref{zetaq}) would give $\alpha_0 = 2.85$, which lacks satisfactory precision. On the other hand, if the negative orders are neglected, the best fit is found for $\alpha_0 =2.2$, now a tolerable value. 
We conclude in this way that stronger finite size effects occur for the case $q < 0$. 

There is in fact a simple phenomenological interpretation of the above result, along the lines of the cascade picture of localization addressed in Ref. [17]. There, a random $\beta$-model approach has been implemented, allowing us to write the multifractal scaling exponents as $\zeta(q) = 2q - \log_a \langle \beta^{-q} \rangle$.
The random parameter $0 < \beta < \infty$ is used to model regions of expanding/collapsing evolution of the wavefunction support, in a discrete iteration defined by the scale compression parameter $a>1$ (i.e., sub-regions with linear sizes $L_0$, $L_0/a$, $L_0/a^2$, ... are successively generated). Essentially, the iteration of a given ``coherent" region which belongs to the wavefunction support is identified to a mapping with local jacobian $\beta$. Therefore, taking $q<0$ in the latter expression for $\zeta(q)$, it is clear that $\beta > 1$ becomes relevant -- and so do expanding fluctuations of the wavefunction -- which we expect will lead to stronger finite size effects.
\begin{figure}[tbph]
\includegraphics[width=7.5cm, height=6.8cm]{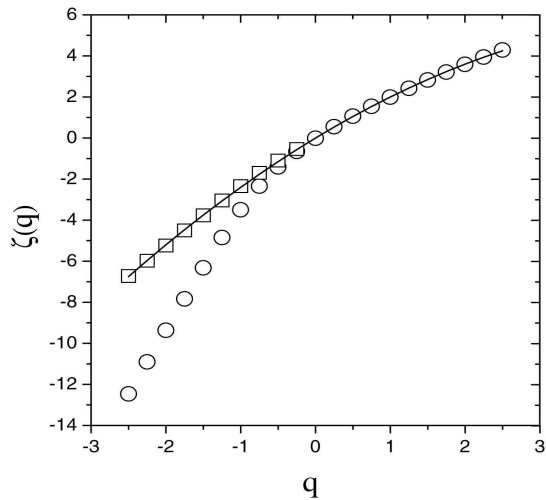}
\caption{The numerical scaling exponents for the probability structure functions (circles), with orders $-2.5 \leq q \leq 2.5$. The solid line corresponds to the theoretical exponents predicted from the parabolic approximation for the multifractal spectrum, with $\alpha_0=2.2$. The squares are the estimated correct values of exponents for negative orders, taking ESS into account.}
\label{fig4}
\end{figure}

Even though the values of $\zeta(q)$ for $q < 0$ strongly deviate from the parabolic approximation predictions, we have verified that the ESS slopes are close to the exact values: taking, for instance, $p=-2.5$ and $q=-1.5$, we get $\eta(-2.5,-1.5)=2.25$, just $4 \%$ off the exact value, obtained from (\ref{eta-ess}), which is $7/3 \simeq 2.33$. This suggests that the numerical results for $\zeta(q)-2q$ are proportional to the exact ones.
Taking this hypothesis into account, we have devised an optimization procedure whereby the solid line (obtained from the parabolic approximation) and the squares (a recovery of more precise multifractal exponents) are found simultaneously in Fig.4. We replace the multifractal exponents for $q < 0$ from their measured values $\zeta(q)$ to $c (\zeta(q)-2q)+2q$. The parameters $c$ and $\alpha_0$ are chosen to be the ones that yield the minimum deviation from the solid line. We find in this way $c=0.23$ and $\alpha_0 = 2.2$. If a more restrict range of $q$'s is considered, say $-1.5 \leq q \leq 1.5$, we find $c= 0.36$ and $\alpha_0=2.27$. Thus, we estimate $\alpha_0 = 2.24 \pm 0.04$ which is in excellent agreement with previous known results. Observe that the only bias in the method is the fact that the multifractal exponents for positive orders were preserved.

In summary, we have numerically established the existence of ESS in the quantum Hall metal-insulator transition for
a small system ($L \simeq 34 \ell$), where strong finite size effects cannot be avoided. We observed that the scaling behavior of structure functions with negative orders is not derived from the parabolic approximation. However, an optimization scheme based on ESS considerations was introduced, which led to a more precise determination of the scaling exponents for negative orders. We found an agreement with known standard results, obtained on their turn from the analysis of much larger systems. We hypothesize that the ESS phenomenon could be used as a valuable tool in order to improve the scaling exponents presently reported in the literature for the quantum Hall effect. A theoretical foundation of ESS is highly desirable, likely through field theory methods, which could allow to check if the hidden scaling behavior found in this work is indeed a general property of metal-insulator transitions.
%\vspace{-0.11cm}

This work has been partially supported by CAPES, FAPERJ and FAPESP.

\end{document}